\documentclass[10pt]{article}
\usepackage[square,authoryear]{natbib}
\usepackage{graphicx}
\usepackage{stackrel}
\usepackage{cancel}

\usepackage{epstopdf,framed}
\usepackage{pdfsync}
\usepackage{amscd}
\usepackage{mathrsfs}

\RequirePackage{color}
\usepackage{amsmath,bm}
\usepackage{amssymb}
\usepackage{amsfonts}
\usepackage{url}

\newtheorem{theorem}{Theorem}[section]

\newtheorem{proposition}[theorem]{Proposition}

\begin{document}
\newtheorem{remark}[theorem]{Remark}

\title{Hamiltonian variational formulation for nonequilibrium  thermodynamics of  simple closed systems\vspace{1cm}}

\date{}
\author{\hspace{-1cm}
\begin{tabular}{cc}
Hiroaki Yoshimura & Fran\c{c}ois Gay-Balmaz
\\   School of Science and Engineering & CNRS, LMD, IPSL
\\  Waseda University & Ecole Normale Sup\'erieure 
\\  Okubo, Shinjuku, Tokyo 169-8555, Japan & 24 Rue Lhomond 75005 Paris, France 
\\ yoshimura@waseda.jp & gaybalma@lmd.ens.fr \\
\end{tabular}\\\\
}
\maketitle
\vspace{-0.3in}

\newcommand{\todo}[1]{\vspace{5 mm}\par \noindent
\framebox{\begin{minipage}[c]{0.45 \textwidth}
\tt #1 \end{minipage}}\vspace{5 mm}\par}

\begin{abstract}
In this paper, we develop a Hamiltonian variational formulation for the nonequilibrium thermodynamics of simple adiabatically closed systems that is an extension of Hamilton's phase space principle in mechanics. We introduce the Hamilton-d'Alembert principle for thermodynamic systems by considering nonlinear nonholonomic constraints of thermodynamic type. In particular, for the case in which the given Lagrangian is degenerate, we construct the Hamiltonian by incorporating the primary constraints via Dirac's theory of constraints. We illustrate our Hamiltonian variational formulation with some examples of systems with friction, with internal matter transfer as well as with chemical reactions.

\end{abstract}

\tableofcontents


\section{Introduction}
As is well known in mechanics, the Euler-Lagrange equations can be obtained from  Hamilton's principle, which has been extensively used in various fields of mechanics as well as in classical field theories such as fluids, electromagnetism, and gravitational fields. Nonequilibrium thermodynamics, however, has not been well established in the context of Hamilton's principle because it was unclear how to incorporate the entropy production due to irreversible processes into the variational principle.  A novel Lagrangian formulation for nonequilibrium thermodynamics extending Hamilton's principle was developed in \cite{GBYo2017a, GBYo2017b} and subsequent works, in which we considered various thermodynamic systems that include irreversible processes such as friction, viscosity, diffusion, heat and mass transfer. 

In this paper, we propose a Hamiltonian variational formulation based on the \textit{Hamilton-d'Alembert principle} which is an extension of Hamilton's phase space principle on the momentum phase space. It arises as a Hamiltonian analogue of the Lagrange-d'Alembert principle on the velocity phase space. In particular, we treat the case of nonlinear nonholonomic constraints of thermodynamic type.

There is however an essential difficulty in the construction of the Hamiltonian variational formulation for thermodynamics. Since the given Lagrangian in thermodynamics is always degenerate, one cannot define a Hamiltonian in a usual way and we need to consider constraints due to the degeneracy of the Lagrangian. To overcome this problem, we consider the following two cases:
\begin{itemize}
\item[(i)] The case in which the given Lagrangian is regular with respect to mechanical variables;
\item[(ii)] The case in which the given Lagrangian is totally degenerate. 
\end{itemize}
In particular, for the second case, we  use Dirac's theory of constraints to construct a Hamiltonian by incorporating the constraints due to the degeneracy of the Lagrangian, and then apply the Hamilton-d'Alembert principle with nonlinear nonholonomic constraints of thermodynamic type.


This paper is organized as follows. In \S\ref{Sect:RevLag}, we review the Lagrangian variational formulation of simple adiabatically closed systems. In \S\ref{Sec:HVM}, we illustrate the abstract setting of the Hamiltonian variational formulation, i.e., the Hamilton-d'Alembert principle for nonholonomic systems with nonlinear constraints of thermodynamic type. In \S\ref{Sect:HamVarForm}, we apply the Hamiltonian variational formulation to the case of simple adiabatically closed systems in which the given Lagrangian is regular with respect to mechanical variables. In \S\ref{Sec:HamFormDegCase}, we consider the general case in which the Lagrangian is degenerate with respect to both mechanical and thermodynamic variables, and employ Dirac's theory of constraints to construct a Hamiltonian and to formulate the equations of motion. We illustrate our theory with thermo-mechanical systems with friction, thermodynamic systems with internal matter transfer as well as thermodynamic systems with chemical reactions.

\section{Review on the Lagrangian variational formulation}\label{Sect:RevLag}
\paragraph{Fundamental setting.} Before starting to discuss the Hamiltonian variational formulation for nonequilibrium thermodynamics, we make a short review on the Lagrangian variational formulation for thermodynamics proposed in \cite{GBYo2017a,GBYo2017b,GBYo2019}.

In this paper, we focus on the case in which the system is {\it simple and adiabatically closed}; namely, the macroscopic thermodynamic property of the system can be represented by one thermodynamic variable, usually an entropy variable, and the system does not exchange heat and matter with the exterior.

Let $Q$ be a $n$-dimensional configuration manifold associated with the mechanical variables of the simple system and let $TQ$ and $T^*Q$ be its tangent and cotangent bundles. Now, suppose that the Lagrangian of the simple thermodynamic system is given  as
\begin{equation}\label{LagThermo}
L: TQ \times \mathbb{R}  \rightarrow \mathbb{R} , \quad (q, v, S) \mapsto L(q, v, S),
\end{equation}
where $q \in Q$ is the mechanical variable, $v \in T_{q}Q$ is the velocity, and $S \in\mathbb{R}$ is the entropy. Assume that the system is subject to external and friction forces respectively given by fiber preserving maps $F^{\rm ext}, F^{\rm fr}:TQ\times \mathbb{R} \rightarrow T^* Q$.

\paragraph{Lagrangian variational formulation.} The Lagrangian variational formulation is given by the Lagrange-d'Alembert principle as follows. Find a curve $(q(t),S(t)) \in Q \times \mathbb{R}$, $t \in [t _1 , t _2 ] \subset \mathbb{R}$ that is critical for the action integral 
\begin{equation}\label{LdA_thermo_simple}
\delta \int_{t _1 }^{ t _2}L(q , \dot q , S)dt +\int_{t_1}^{t_2}\left\langle F^{\rm ext}(q, \dot q, S), \delta q\right\rangle dt =0,
\end{equation}
for all variations $ \delta q(t) $ and $\delta S(t)$, which are subject to the \textit{variational constraint}
\begin{equation}\label{CV_simple} 
\frac{\partial L}{\partial S}(q, \dot q, S)\delta S= \left\langle F^{\rm fr}(q , \dot q , S),\delta q \right\rangle,
\end{equation}
with $ \delta q(t_1)=\delta q(t_2)=0$, and the curve is subject to the \textit{phenomenological  constraint}
\begin{equation}\label{CK_simple} 
\frac{\partial L}{\partial S}(q, \dot q, S)\dot S  = \left\langle F^{\rm fr}(q, \dot q, S) , \dot q \right\rangle,
\end{equation}
where $ \dot q=\frac{dq}{dt}$ and $\dot S=\frac{dS}{dt}$.
Taking the variations of the action integral in \eqref{LdA_thermo_simple} and using the variational constraint \eqref{CV_simple}, we get the system of evolution equations for the simple thermodynamic system together with the phenomenological constraint \eqref{CK_simple} as
\begin{equation}\label{simple_systems} 
\left\{
\begin{array}{l}
\displaystyle\vspace{0.2cm}\frac{d}{dt}\frac{\partial L}{\partial \dot q}- \frac{\partial L}{\partial q}=  F^{\rm fr}+F^{\rm ext},\\
\displaystyle\frac{\partial L}{\partial S}\dot S= \left\langle F^{\rm fr}, \dot q \right\rangle.
\end{array} \right.
\end{equation} 
Along the solution curve $(q(t),S(t))$ of \eqref{simple_systems}, the first law of thermodynamics is verified as 
\[
\frac{d}{dt} E= \left\langle F^{\rm ext}, \dot q \right\rangle,
\]
where the energy is given by
$
E(q,\dot q, S)= \left\langle \frac{\partial L}{\partial \dot q}(q, \dot q, S), \dot q \right\rangle - L(q, \dot q, S).
$
Note that the temperature of the system $T$ is defined by $T:=- \frac{\partial L}{\partial S}$, while the friction force is given by $F^{\rm fr}(q, v, S)=- r (q, S) v$, where $ r(q, S)=[r_{ij}(q,S)], i,j=1,...,n$ is the phenomenological coefficient,  determined experimentally and the symmetric part of the matrix $r=[r_{ij}]$ is positive semi-definite. Hence it follows from the second equation of \eqref{simple_systems} that the internal entropy production $I= \frac{1}{T}\left<rv,v\right>$ is always positive, consistently with the second law.

\section{Hamilton's variational formulation with nonholonomic constraints}\label{Sec:HVM}

\subsection{Hamilton-d'Alembert principle in mechanics}
First, let us review Hamilton's variational formulation in nonholonomic mechanical systems, on which linear nonholonomic constraints are imposed. 

Given a {\it hyperregular} Lagrangian $L=L(q,v)$ on $TQ$ and an external force  $F^{\rm ext}:TQ \rightarrow T^* Q$, we can define
 a hyperregular Hamiltonian $H=H(q,p)$ on the cotangent bundle (momentum phase space) $ T ^\ast Q$, with $2n$ canonical coordinates $(q^{i},p_{i}),i=1,...,n$ for $(q,p) \in T^{\ast}Q$, by $H=E \circ (\mathbb{F}L)^{-1}$, where $E(q,v):=\left<\mathbb{F}L(q,v),v\right>-L(q,v)$ is the energy and $ \mathbb{F} L:TQ \rightarrow T^*Q$ is the fiber derivative of $L$. 
 
Consider nonholonomic constraints given by a distribution $\Delta_{Q}$ on $Q$ written as
\[
\Delta_{Q}(q):=\{ (q,v)\in T_{q}Q \mid \left<\omega^{r}(q), v\right>=0,\;r=1,...,m<n\},
\]
where $\omega^{r}=\sum_{i=1}^{n}\omega^{r}_{i}(q)dq^{i}$ are given $m$ one-forms on $Q$.

Define an external force field $\mathcal{F}^{\rm ext}: T^{\ast}Q \to T^{\ast}Q$ by $\mathcal{F}^{\rm ext}:={F}^{\rm ext} \circ (\mathbb{F}L)^{-1}$.
Then, the {\it Hamilton-d'Alembert principle} is given by the critical condition of the action functional as follows (see \cite{YoMa2006b}):
\begin{equation*}
\begin{split}
&\delta \int_{t_{0}}^{t_{1}} \Big[\! \left\langle p, \dot q \right\rangle -H(q, p) \Big] dt +\int_{t_{0}}^{t_{1}} \left<\mathcal{F}^{\rm ext}(q,p), \delta{q} \right>dt=0,
\end{split}
\end{equation*}

where $\dot q(t)  \in \Delta _Q(q(t))$, and for variations $ \delta q(t), \delta p(t)$ such that $ \delta q(t)  \in \Delta _Q(q(t)) $ and with $\delta{q}(t_{0})=\delta{q}(t_{1})=0$.
This principle yields the {\it Hamilton-d'Alembert equations} for nonholonomic mechanics:
\[
\dot q = \frac{\partial H}{\partial p}\in \Delta _Q , \quad \dot p +\frac{\partial H}{\partial q}-\mathcal{F}^{\rm ext}\in \Delta ^\circ_Q.
\]
In local coordinates, we get 
\begin{equation*}
\omega^{r}_{i}(q)\frac{\partial H}{\partial p_{i}} =0,\quad
\dot{p}_{i}+ \frac{\partial H}{\partial q^{i}} =\mathcal{F}^{\rm ext}_{i}(q,p)+\lambda_{r}\omega^{r}_{i}(q,p),
\end{equation*}
where $\lambda_{r}$ are $m$ Lagrange multipliers.
As to the geometric formulation of Hamiltonian systems with {\it linear} nonholonomic constraints; see \cite{BaSn1993}.

\subsection{Nonlinear constraints of the thermodynamic type}\label{subsec_32}

We present an abstract setting for the Hamiltonian variational formulation of nonequilibrium thermodynamic systems with nonlinear nonholonomic constraints of thermodynamic type (see, \cite{GBYo2017a}). In this case, the  variational constraint $C_{V} \subset T Q \oplus T Q $ and kinematic constraint $C_{K} \subset TQ$ are related as
\begin{equation}\label{def_CK}
C_{K}:=\{ (q,v) \in  T Q \mid (q,v)  \in C_{V}(q,v)\}.
\end{equation} 
If condition $(q,v, \delta q) \in C_{V}$ is locally given by $ A_i ^r(q,v)    \delta q^i =0,\; r=1,...,m,\,i=1,...,n$, then $C_{K}$ is locally given by $ A_i ^r (q,v)   v^i=0 $. We define $\mathscr{C} _V \subset T^*Q \oplus  TQ$ as
\begin{equation}\label{def_CCCV} 
\mathscr{C} _V(q,p):= C_V(q,v),
\end{equation} 
where we assume that the right hand side does not depend on the choice of $v$ such that $ \frac{\partial L}{\partial v}(q,v)=p$. This assumption is of course satisfied when the Lagrangian is hyperregular, but also in all the cases of interests in thermodynamics although the Lagrangian is necessarily degenerate in that case.
Locally, the condition $(q,p, \delta q) \in \mathscr{C}_{V}$ is given by $ \mathcal{A}_i ^r(q,p)    \delta q^i =0,\; r=1,...,m,\,i=1,...,n$.

\subsection{The Hamilton-d'Alembert principle for nonholonomic systems with nonlinear constraints of thermodynamic type}\label{Thm:AbstHDAP}
Associated with the Hamiltonian $H$ on $T^{\ast}Q$,
the Hamiltonian variational formulation of nonholonomic systems with nonlinear constraints of thermodynamic type is given by the following theorem. 

\begin{theorem}\label{HamDLAP_NH}
A curve $(q(t),p(t))\in T^{\ast}Q$ satisfies the Hamiltonian equations of motion 
\begin{equation}\label{NNHamEqn}
\dot{q}= \frac{\partial H}{\partial p} \in \mathscr{C}_V(q,p),\;\;
\dot{p}+ \frac{\partial H}{\partial p} \!-\! \mathcal{F}^{\rm ext}(q,p)\in \mathscr{C}^{\circ}_{V}(q,p),
\end{equation} 
%
if it is a critical curve of the action integral:
\begin{equation*}
\delta \int_{t_1}^{t_2}\Big[\! \left\langle p, \dot{q}\right\rangle -H(q,p)\Big] dt+\int_{t_{0}}^{t_{1}} \left<\mathcal{F}^{\rm ext}(q,p), \delta{q} \right>dt=0,
\end{equation*}
for variations $ \delta{q}, \delta{p}$ such that $ \delta{q} \in \mathscr{C}_V(q,p)$ and $ \delta{q}(t_1)= \delta{q}(t_2)=0$, and with the constraint $\dot{q} \in \mathscr{C}_V(q, p)$. 
\end{theorem}
The variational formulation in Theorem \ref{HamDLAP_NH} is called the \textit{Hamilton-d'Alembert principle} and the equations of motion in \eqref{NNHamEqn} are the Hamilton-d'Alembert equations.

In local coordinates, the equations \eqref{NNHamEqn} are given as
\begin{equation*}
\left\{
\begin{array}{l}
\displaystyle\vspace{0.2cm}\dot{q}^{i}=\frac{\partial H}{\partial p_{i}},\quad
\dot{p}+ \frac{\partial H}{\partial q} =\mathcal{F}^{\rm ex}_{i}(q,p)+\lambda_{r}\mathcal{A}^{r}_{i}(q,p),\\
\displaystyle\vspace{0.2cm}\mathcal{A}^{r}_{i}(q,p)
\frac{\partial H}{\partial p_{i}} =0,\; \; r=1,...,m,\,i=1,...,n.
\end{array} \right.
\end{equation*} 

\section{Hamiltonian variational formulation of nonequilibrium thermodynamics}\label{Sect:HamVarForm}

\subsection{Simple adiabatically closed systems with friction}

\paragraph{Nonlinear constraints of thermodynamic type.}
Consider the case of a simple closed thermodynamic system with a Lagrangian $L=L(q,v,S)$ defined on $TQ \times \mathbb{R}$ as in \eqref{LagThermo}.

Assume here that $L$ is hyperregular with respect to the mechanical part $(q,v) \in TQ$ so that the partial Legendre transform $\mathbb{F}L_S: TQ   \rightarrow T^* Q$  defined, for each fixed $S \in \mathbb{R}$, as 
\begin{equation*}
(q,v) \mapsto \left( q, \frac{\partial L}{\partial v}(q,v,S) \right)
\end{equation*} 
is a diffeomorphism.
Hence we can define a \textit{Hamiltonian} $H=H(q,p,S)$ on $T^* Q \times \mathbb{R}$ by
\begin{equation*}
H(q,p,S)=\left\langle p, v \right\rangle -L(q, v, S),
\end{equation*} 
where $v$ is uniquely determined from $(q,p,S)$ by the condition $\frac{\partial L}{\partial v}(q, v, S)=p$. Let $\mathcal{F}^{\rm ext}:T^{\ast}Q \times \mathbb{R}  \rightarrow T^{\ast}Q$ and $\mathcal{F}^{\rm fr}:T^{\ast}Q \times \mathbb{R}  \rightarrow T^{\ast}Q$ be the external and friction forces, each of which is defined such that $\mathcal{F}^{\rm ext}\circ \mathbb{F}L_{S}=F^{\rm ext}$ and $\mathcal{F}^{\rm fr}\circ \mathbb{F}L_{S}=F^{\rm fr}$. Here $Q$ is the configuration manifold of the mechanical variables $q$ of the system, and $ \mathbb{R}  $ denotes the space of the thermodynamic variable. Let us further introduce the thermodynamic configuration manifold $\mathcal{Q} :=  Q\times \mathbb{R}$. Then, the variational constraint is
\begin{equation}\label{C_V_thermo}
\begin{split}
\mathscr{C} _V&= \biggl\{(q,S, p, \Lambda,\delta q, \delta S) \in T^{\ast} \mathcal{Q} \times _ \mathcal{Q} T \mathcal{Q}  \; \biggm\vert \; \biggr.\\
&\qquad\qquad \biggl.-\frac{\partial H}{\partial S}(q, p , S)\delta S= \left\langle \mathcal{F}^{\rm fr}(q , p , S),\delta q \right\rangle\biggr\},
\end{split}
\end{equation} 
where $(q,S)\in \mathcal{Q}$, $(p, \Lambda) \in T^{\ast}_{(q,S)} \mathcal{Q}$, and $(\delta q, \delta S) \in T_{(q,S)} \mathcal{Q}$. Note that the temperature of the system is defined by $T:=\frac{\partial H}{\partial S}(q, p , S)$.
By hypothesis $\frac{\partial H}{\partial S}(q, p , S)\neq 0$, the space $\mathscr{C}_V$ is a submanifold of $T^{\ast} \mathcal{Q} \oplus T \mathcal{Q}$ of codimension one.

For each $(q,S, p, \Lambda) \in T^{\ast}\mathcal{Q} $, the annihilator $\mathscr{C} ^\circ_V(q,S,p, \Lambda)$ of the variational constraint $\mathscr{C} _V(q,S, p, \Lambda)$ reads
\begin{equation*}
\left \{ (q,S, \alpha , \mathcal{T}   ) \in T^*_{(q,S)} \mathcal{Q} \; \biggm\vert \; \right.\\
\left.\alpha  \frac{\partial H}{\partial S}(q, p, S)=  \mathcal{T}    F^{\rm fr}(q, p, S) \right\}.
\end{equation*} 

For simple adiabatically closed systems, the Hamilton-d'Alembert principle is given by the following proposition.
\begin{proposition}
A curve $(q(t),p(t),S(t))$ on $T^{\ast}Q \times \mathbb{R}$ satisfies the Hamilton-d'Alembert equations:
\begin{equation}\label{thermo_mech_equations}
\begin{split}
\left\{ 
\begin{array}{l}
\displaystyle\vspace{0.2cm}\dot{q}= \frac{\partial H}{\partial p},\\
\displaystyle\vspace{0.2cm}\dot{p}+ \frac{\partial H}{\partial q}= \mathcal{F}^{\rm ext}(q , p, S) +  \mathcal{F}^{\rm fr}(q, p, S),\\
-\displaystyle\frac{\partial H}{\partial S}\dot S= \left\langle \mathcal{F}^{\rm fr} (q , p, S) , \dot q\right\rangle,
\end{array} 
\right.
\end{split}
\end{equation}
if it is a critical curve of the action integral  
\begin{equation*}
\begin{split}
&\delta \int_{t_1}^{t_2} \!\Big[\left\langle p, \dot q \right\rangle - H (q,p,S )\Big] dt \!+\!\int_{t_{0}}^{t_{1}} \!\left<\mathcal{F}^{\rm ext}(q,p,S), \delta{q} \right>dt=0
\end{split}
\end{equation*} 
for $(\delta q, \delta p, \delta S)$ satisfying the variational constraint
\begin{equation*}
-\frac{\partial H}{\partial S}(q,p,S) \delta  S= \left\langle \mathcal{F}^{\rm fr}(q,p,S),\delta q \right\rangle 
\end{equation*}
with $\delta q(t_1 )= \delta q(t _2 )=0$, and where the curve $(q(t),p(t),S(t))$ is  subject to the phenomenological constraint
\begin{equation*}
-\frac{\partial H}{\partial S}(q,p,S) \dot S= \left\langle \mathcal{F}^{\rm fr}(q,p,S),\dot q \right\rangle.
\end{equation*}
\end{proposition}
\paragraph{The first law.} We can easily check that along the solution curve $(q(t),p(t),S(t))$ of \eqref{thermo_mech_equations}, we have the energy balance
$$
\frac{d}{dt} H(q(t), p(t), S(t))= P_W^{\rm ext}, 
$$
where $P_W^{\rm ext}=\left<\mathcal{F}^{\rm ext}(q,p,S),\dot{q}\right>$ is the mechanical power.

\paragraph{Entropy production.} Recall that the temperature is given by $T=\frac{\partial H}{\partial S}$, which is assumed to be positive. From the third equation in \eqref{thermo_mech_equations}, it follows
\[
T \dot S=- \left\langle \mathcal{F}^{\rm fr}(q, p, S), \dot q\right\rangle .
\]
From  the second law we must have $\left\langle \mathcal{F}^{\rm fr}(q, p, S), \dot q \right\rangle \leq 0$ for all $(q, p, S) \in T^{\ast}Q \times \mathbb{R}  $, i.e. the friction force $\mathcal{F}^{\rm fr}$ is dissipative. We can hence assume the phenomenological relation $\mathcal{F}^{\rm fr}_{i}=-r_{ij}\frac{\partial H}{\partial p_j}$, where $r_{ij}$, $i,j=1,...,n$ are functions of $(q,S)$ and where the symmetric part of the matrix $r =[r_{ij}]$ is positive semi-definite.

\subsection{Example: one piston-cylinder system}

Consider a simple adiabatically closed system made from a piston-cylinder arrangement with  an ideal gas. 
The system is energized by an external force $\mathcal{F}^{\rm ext}(q, p, S)$ and the state of the system can be described by $(q, p, S)$.


The Hamiltonian is $H(q,p, S)=\frac{1}{2} \frac{p^{2}}{m} +U(q,S)$, in which $m$ is the mass of the piston, $U(q,S):=\mathsf{U}(S,V=\alpha q,N_0)$, with $\mathsf{U}(S,V,N)$ the internal energy of the gas, $N_0$ is the constant number of moles, $V=\alpha q$ is the volume, and $\alpha$ is the constant sectional area of the cylinder. The friction force is $\mathcal{F}^{\rm fr}(q, p, S)=- r (q, S) \frac{p}{m}$, where $ r(q, S)\geq 0$ is the phenomenological coefficient,  determined experimentally.

From \eqref{thermo_mech_equations}, the equations of motion for the piston-cylinder system are obtained as
\begin{equation*}
\dot q=m^{-1}p,\;\;
\dot{p} =\textsf{p}(q,S)\alpha+\mathcal{F}^{\rm ext}- r (q, S)\dot q,\;\;
\dot S= \frac{r m^{-2}}{T} p^2,
\end{equation*}
where the pressure is given by $\textsf{p}=- \frac{\partial \mathsf{U}}{\partial V}=- \alpha^{-1}\frac{\partial \mathsf{U}}{\partial q}$ and the temperature is defined as $T:=\frac{\partial U}{\partial S}(q,S)$.  From the third equation, the internal entropy production is positive for all time $t$, consistently with the second law. We can also verify the first law of energy balance since $\frac{d}{dt} H(q,p,S)=\mathcal{F}^{\rm ext}\dot q$ holds along the solution curve $(q(t),p(t),S(t))$.


\subsection{Simple systems with internal mass transfer}

Consider a thermodynamic system with an internal diffusion process, which typically appears in biological systems where many chemical processes accompany the mass transfer of chemical species through membranes; see \cite{OsPeKa1973}. Suppose that the system has $K$ compartments with common boundaries consisting of walls (or membranes), through which matter is exchanged by diffusion. 
Assume that the system involves mechanical state variables $(q,p)$, friction and external forces $\mathcal{F}^{\rm fr}$, $\mathcal{F}^{\rm ext}$, and a single species with number of moles $N_k$ in the $k$-th compartment, $k=1,...,K$. We suppose that the system has one single entropy $S$ that represents the macroscopic thermodynamic state of the system, attributed to all the compartments.
For each compartment $k=1,...,K$, we get the mole balance equation as
$
\frac{d}{dt} N _k = \sum_{\ell=1}^K \mathcal{J} ^{\ell \rightarrow k},
$
where $\mathcal{J} ^{\ell \rightarrow k}=- \mathcal{J} ^{k \rightarrow \ell}$ indicates the molar flow rate from compartment $\ell$ to compartment $k$ due to diffusion of the species. The Hamiltonian of the system is
\[
\begin{split}
&H: T^{\ast}Q\times\mathbb{R} \times \mathbb{R}^K  \rightarrow \mathbb{R} , \\
& \qquad \left(q, p,S, N_1,...,N_K\right) \mapsto H\left(q, p,S, N_1,...,N_K\right).
\end{split}
\]

\paragraph{Thermodynamic displacements.} Let us introduce the \textit{thermodynamic displacements} $W^k$, $k=1,...,K$, which play an essential role in our variational formulation. More generally, the thermodynamic displacement associated with an irreversible process can be defined as the primitive in time of the thermodynamic force (or affinity) of the process. 
When we consider matter transfer, $\dot W^k$ becomes the chemical potential $\mu^{k}$ associated with $N_k$.

\paragraph{The Hamiltonian variational formulation for simple systems with diffusions.} The Hamilton-d'Alembert principle holds as follows. Find the curves $q(t)$, $p(t)$, $S(t)$, $W^k(t)$, $N_k(t)$, which are critical for the \textit{variational condition}
\begin{equation*}
\begin{split}
\delta \int_{t _1 }^{ t _2}& \!\Big[ \left<p,\dot{q}\right>-H\left(q, p, S, N_1,...,N_K\right)+ \dot W^kN_k\Big] {\rm d}t \\
& \qquad \qquad + \int_{t_1}^{t_2} \left<\!\mathcal{F}^{\rm ext }, \delta q\right>\,{\rm d}t=0, 
\end{split}
\end{equation*}
subject to the \textit{phenomenological constraint}
\begin{equation*}
-\frac{\partial H}{\partial S}\dot S  =  \left<\mathcal{F}^{\rm fr}, \dot q\right>   + \sum_{k,\ell=1}^{K}  \mathcal{J}^{\ell \rightarrow k} \dot W^k,
\end{equation*}
and for variations subject to the \textit{variational constraint}
\begin{equation*}
-\frac{\partial H}{\partial S}\delta S  = \left< \mathcal{F}^{\rm fr},\delta q \right> + \sum_{k,\ell=1}^{K}  \mathcal{J}^{\ell \rightarrow k} \delta W ^k,
\end{equation*}
with $\delta q(t_1)=\delta q(t_2)=0$ and $ \delta W^k(t_1)=\delta W^k(t_2)=0$.
From this we obtain the system of evolution equations for the curves $q(t)$, $p(t)$, $S(t)$, $W^k(t)$ and $N_k(t)$ as:
\begin{equation}\label{simple_systems_matter} 
\left\{
\begin{array}{l}
\displaystyle\vspace{0.1cm}\dot{q}= \frac{\partial H}{\partial p},\;\; \dot{p}=-\frac{\partial H}{\partial q}  +\mathcal{F}^{\rm fr}+\mathcal{F}^{\rm ext }, \\
\displaystyle\vspace{0.1cm} \dot W^{k}=\frac{\partial H}{\partial N_k},\quad k=1,...,K,\\
\displaystyle\vspace{0.1cm}\dot N _k = \sum_{\ell=1}^K \mathcal{J} ^{\ell \rightarrow k}, \quad k=1,...,K,\\
\displaystyle-\frac{\partial H}{\partial S}\dot S=  \left<\mathcal{F}^{\rm fr},\dot q \right> + \sum_{k<\ell} \mathcal{J}^{\ell\rightarrow k}\left(\frac{\partial H}{\partial N_k} - \frac{\partial H}{\partial N_\ell}\right) .
\end{array} \right.
\end{equation} 

\paragraph{The first law.} By taking the time derivative of the Hamiltonian $H\left(q, p,S, N_1,...,N_K\right)$ along the solution curve of  \eqref{simple_systems_matter}. We get
\[
\begin{split}
\frac{d}{dt}H
&= \mathcal{F}^{\rm ext}_i\dot q ^i= P^{\rm ext}_W,
\end{split}
\]
where $P_W^{\rm ext}$ is the mechanical power associated with $\mathcal{F}^{\rm ext}$ that is done on the system, consistently with the first law.

\medskip

Associated with the given Hamiltonian $H$, recall that by definition the temperature of the system and the chemical potentials of each compartment are given by
\[
T:=\frac{\partial H}{\partial S}\qquad\text{and}\qquad \mu^k:=\frac{\partial H}{\partial N_k},\;\; k=1,...,K.
\]
From the last equation in \eqref{simple_systems_matter}, the rate of entropy production of the system becomes
\[
\dot S=-\frac{1}{T} \left<\mathcal{F}^{\rm fr}, \dot{q} \right>- \frac{1}{T}\sum_{k<\ell}
\mathcal{J} ^{k \rightarrow \ell} (\mu^k -\mu^\ell),
\]
where the two terms on the right-hand side correspond, respectively, to the rate of entropy production due to mechanical friction and to matter transfer.
The second law suggests the phenomenological relations
\[
\mathcal{F}^{\rm fr}_{i}=-\lambda_{ij}\frac{\partial H}{\partial p_j}\qquad\text{and}\qquad \mathcal{J} ^{k \rightarrow \ell}=G^{k\ell} (\mu ^k-\mu^\ell),
\]
where $\lambda_{ij}$, $i,j=1,...,n$ and $G^{k\ell}$, $k,\ell=1,...,K$ are functions of the state variables, with the symmetric part of the matrix $\lambda_{ij}$ positive semi-definite and with $G^{k\ell}\geq 0$, for all $k,\ell$.

\section{Hamiltonian formulations for the degenerate cases}\label{Sec:HamFormDegCase}
\subsection{Dirac's theory of constraints}
\paragraph{Motivations for the degenerate cases.}
We have assumed that the given Lagrangian $L=L(q,v,S)$ on $TQ\times \mathbb{R}$ is hyperregular with respect to the mechanical state variables $(q,v) \in TQ$, although the Lagrangian $L=L(q,v,S)$ itself is already degenerate with respect to the thermodynamic variable $S$. Further, in general, there exists the nontrivial case in which the given Lagrangian $L=L(q,v,S)$ is totally degenerate with respect to both the mechanical variables as well as the thermodynamic variable. Such a case will appear in the example of chemical reaction dynamics.
In such situations, we cannot define a Hamiltonian on $T^{\ast}Q \times \mathbb{R}$ as done in \S\ref{Sect:HamVarForm} but in order to go over to the Hamiltonian side, we need to employ {\it Dirac's theory of constraints}; see \cite{Dirac1964} and also \cite{YoMa2007} in conjunction with Dirac structures.
\paragraph{Primary constraints.}
Let us consider the abstract setting for Dirac's theory of constraints. Let $L=L(q,v)$ be a given Lagrangian, possibly degenerate, in which case the determinant of the Hessian matrix of $L(q,v)$ is zero, i.e.,
\begin{equation*}
\textrm{det}
\left[
\frac{\partial^{2} L}{\partial v^{i} \partial v^{j}}
\right]=0.
\end{equation*}
Define the constraint set  $P \subset T^{\ast}Q$ as the image of the Legendre transform $\mathbb{F}L: TQ \to T^{\ast}Q; (q,v) \mapsto \mathbb{F}L(q,v)=\left(q, \frac{\partial L}{\partial v}\right)$, i.e.,  
$P=\mathbb{F}L(TQ)$.
We assume that $P$ is a $2n-m$ submanifold of $T^*Q$, $1 \le m < 2n$, locally written as
\begin{equation}\label{PrimaryConstraints}
P=\left\{ (q,p) \in T^{\ast}Q \mid  \phi_A(q,p)=0,  \; A=1,...,m \right \},
\end{equation} 
where $\phi_A(q,p),\,A=1,...,m,$ are smooth functions on $T^\ast{Q}$. The constraints  $\phi_A(q,p)=0$ in equation \eqref{PrimaryConstraints} are the so-called {\it primary constraints} (see \cite{Dirac1964}). 

\paragraph{Constrained and total Hamiltonians.}
Define the generalized energy $E: TQ \oplus T^\ast Q \to \mathbb{R}$ by 
$E(q,v,p):=\left<p,v\right>-L(q,v)$
and consider the graph $\mathscr{G}\subset TQ \oplus T^\ast Q$ of $ \mathbb{F}L$
\begin{equation*}
\begin{split}
\mathscr{G}:=\left\{ (q,v,p) \in TQ \oplus T^\ast Q   \bigm\vert   p=  \mathbb{F}L(q,v)\right\}.
\end{split}
\end{equation*}
Note that $ \frac{\partial E}{\partial v}(q,v,p)=0$ for all $(q,v,p) \in \mathscr{G}$; see \cite{TuUr1999,YoMa2006b}.


Assume that there is a smooth function $\widetilde H: U \subset T^*Q \rightarrow \mathbb{R} $ defined on an open subset $U$ containing $P$, such that the condition
\begin{equation}\label{ConHam}
E(q,v,p)=\widetilde H(q,p),
\end{equation}
holds for all $(q,v,p) \in \mathscr{G}$. Such a constrained Hamiltonian $\widetilde{H}$, when it exists, is not unique as we can add to it any smooth function on $U$ which vanishes on $P$. We refer to  \cite{Mar2008} for details, as well as for conditions ensuring the existence of $\widetilde{H}$. As we will see, in our example the construction of such a function $\widetilde{H}$ is natural.

\paragraph{Hamiltonian variational formulation for degenerate systems.}
We develop here an abstract setting for the variational formulation of degenerate systems by using the total Hamiltonian.

Now we assume that the system has nonlinear nonholonomic constraints of thermodynamic type given in \eqref{def_CK} and with an external force $\mathcal{F}^{\rm ext}(q,p)$.

Associated with the function in \eqref{ConHam}, we define the total Hamiltonian on $U \times \mathbb{R}^{m} \subset T^*Q \times \mathbb{R} ^m$ by
\begin{equation}\label{TotHam}
H_{T}(q,p,\lambda)=\widetilde{H}(q,p) + \sum_{A=1}^{m}\lambda^{A} {\phi}_{A}(q,p).
\end{equation}
Such a total Hamiltonian was first considered by \cite{Dirac1964}.
Then, we  apply Theorem \ref{HamDLAP_NH} to the degenerate case with the total Hamiltonian:  Find a curve $(q(t), p(t),\lambda(t))\in U\times \mathbb{R}^{m}$ that is critical for the action integral:
\begin{equation*}
\delta \int_{t_1}^{t_2}\!\Big[\! \left\langle p, \dot{q}\right\rangle - H_{T}(q,p,\lambda)\Big] dt+\int_{t_{0}}^{t_{1}} \left<\mathcal{F}^{\rm ext}(q,p), \delta{q} \right>dt=0,
\end{equation*}
for  arbitrary variations $\delta{\lambda}(t)$ and for variations $ \delta{q}(t), \delta{p}(t)$ such that $ \delta{q} \in \mathscr{C}_V(q,p)$ together with $ \delta{q}(t_1)= \delta{q}(t_2)=0$, where the curves $q(t),p(t)$ are also subject to the constraints $\dot{q} \in \mathscr{C}_V(q, p)$. 

By direct computations, we obtain the set of evolution equations for the curves $q(t)$, $p(t)$, and $\lambda_{A}(t)$ as:
\begin{equation}\label{HamDAEqn_NNH} 
\left\{
\begin{array}{l}
\displaystyle\vspace{0.3cm}
\dot{q}= \frac{\partial  \widetilde{H}}{\partial p} + \lambda^{A}\frac{\partial \phi_{A}}{\partial p}\in \mathscr{C}_{V}(q,p),\\
\displaystyle\vspace{0.3cm} \dot{p}+\frac{\partial \widetilde{H}}{\partial q} + \lambda^{A}\frac{\partial \phi_{A}}{\partial q} 
-\mathcal{F}^{\rm ext} \in \mathscr{C}^{\circ}_{V}(q,p), \quad\\
\displaystyle\vspace{0.2cm}\phi_{A}(q,p)=0,\quad A=1,...,m.
\end{array}\right.
\end{equation}

\subsection{Chemical reaction dynamics}
\paragraph{Setting for chemical reactions.} As a typical example of a degenerate system, we consider the case of several chemical species $I=1,...,R$ undergoing $a=1,...,r$ chemical reactions among them, which are given by
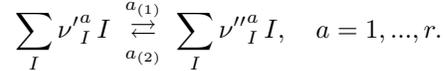
\[
\sum_I {\nu '}_{I}^a\,I\; \mathrel{\mathop{\rightleftarrows}^{a _{(1)}}_{a _{(2)}}}  \; \sum_I{\nu ''}^a_{I}\, I, \quad a=1,...,r.
\]
In the above, $a_{(1)}$ and $a_{(2)}$ indicate respectively the forward and backward reactions associated with the reaction $a$, and also ${\nu''}^a _{I}$, ${\nu '}^a_I$ respectively the forward and backward stoichiometric coefficients for the species $I$ in the reaction $a$. Note that the mass conservation, i.e., \textit{Lavoisier's law} holds as
\[
\sum_I m_I\nu^a_{I}=0, \quad \text{$a=1,...,r$},
\]
where $\nu^a _{I}:= {\nu ''}^a_{I}- {\nu'}^a _{I}$ and $m_I$ is the molecular mass of species $I$. The affinity of reaction $a$ is defined as 
$$
\mathcal{A} ^a= - \sum_I\nu^a _{I} \mu ^I, \;\; a=1,...,r, 
$$
where $\mu^I$ is the chemical potential of $I$. Define the thermodynamic displacements $W^I$ and $\nu^a$  such that
$
\dot W^I=\mu^I
$
and
$
 \dot\nu^a=-\mathcal{A}^a.
$
For each reaction $a$, the {\it thermodynamic flux} $J_a$, dual to the affinity $\mathcal{A}^a$, can be defined by the {\it time derivative  of extent of reaction}, i.e., the {\it reaction rate}.
\paragraph{The Hamiltonian formulation for chemical reactions.}
Suppose that the system of chemical reactions has no external forces or exchange of heat and matter with the exterior and is therefore isolate. We assume that the volume is constant. Let $U(q)=U(N_1,...,N_R, S)$ be an internal energy on the thermodynamic configuration space $ \mathcal{Q}=\mathbb{R}^{R} \times \mathbb{R}$, where $q=(N_1,...,N_R, S) \in  \mathcal{Q}$ and where $N_{I}$ denotes the number of moles of each species $I=1,...,R$ and $S$ denotes the entropy of the system. 

Now we define the Lagrangian $L$ on $T \mathcal{Q}$ by the internal energy as $L(q,v):=-U(q)=-U(N_1,...,N_R, S)$ and it is apparent that $L$ is degenerate since it does not depend on the velocity variables $v=(v_{N_{1}},...,v_{N_{R}},v_{S}) \in T_{q} \mathcal{Q}$. Therefore, we cannot define the Hamiltonian on $T^{\ast} \mathcal{Q}$ in a usual way and we shall employ Dirac's theory of constraints. To do this, we introduce the momentum variables $p=(p_{N_{1}},...,p_{N_{R}},p_{S}) \in T_{q}^{\ast} \mathcal{Q}$, dual to $v_{q}=(v_{N_{1}},...,v_{N_{R}},v_{S})$. Now  the primary constraint  $P=\{ (q,p)\in T^{\ast} \mathcal{Q} \mid (q,p)=\mathbb{F}L(q,v)\}$ may be locally given, see  \eqref{PrimaryConstraints},  by the functions $\phi_{N_{1}},...,\phi_{N_{R}},\phi_{S}$ on $T^{\ast} \mathcal{Q}$ as
\begin{equation*}
P=\{ \phi_{N_{1}}=p_{N_{1}}=0,...,\phi_{N_{R}}=p_{N_{R}}=0, \phi_{S}=p_{S}=0 \}.
\end{equation*}
From the expression of the generalized energy $E(q,v,p)=\left<p,v\right>-L(q,v)= \left\langle p,v \right\rangle + U(q)$ it easily follows that $\widetilde{H}(q,p)=U(q)$ satisfies condition \eqref{ConHam}. 
From equation \eqref{TotHam}, the total Hamiltonian is given by
\begin{equation*}
\begin{split}
H_{T}(q,p,\lambda)
&=U(N_1,...,N_R, S)+\sum_{I=1}^{R}p_{N_{I}}\lambda^{N_{I}}+p_{S}\lambda^{S}.
\end{split}
\end{equation*}
Thus, the Hamiltonian variational formulation for chemical reaction dynamics is given as: Find the curves $S(t)$, $N_I(t)$, $p_{S}(t)$, $p_{N_{I}}(t)$, $\lambda^A(t)$, $I=1,...,R$, $A=N_{1},...,N_{R}, S$, which are critical for the variational condition
\begin{equation}\label{al_GLdA_thermo_chem} 
\begin{split}
&\!\delta \!\int_{ t _1 }^{ t _2 } \! \left[p_{S}\dot{S}+p_{N_{I}}\dot{N}^{I}\!\! -\!H_{T}(N_I,S,\lambda)+\dot{W}^{I}N_{I} \right]\!dt  =0,
\end{split}
\end{equation}
subject to the \textit{phenomenological and chemical constraints}
\begin{equation}\label{al_GNonholonomic_Constraints1_chem} 
-\frac{\partial H}{\partial S}\dot S   = J_a \dot{\nu}^a\quad\text{and}\quad\dot{\nu}^a= \nu^a_I \dot{W}^I,\quad a=1,...,r,
\end{equation}
and for variations subject to the \textit{variational constraints}
\begin{equation}\label{al_GVariational_Constraints_chem} 
-\frac{\partial H}{\partial S}\delta S= J_a \delta{\nu}^a \quad\text{and}\quad \delta{\nu}^a= \nu^a_I \delta W^I,\quad a=1,...,r,
\end{equation}
with $ \delta W ^I ( t _1)=\delta W^I(t_2)=0$, $I=1,...,R$.

Form the variational formulation \eqref{al_GLdA_thermo_chem}--\eqref{al_GVariational_Constraints_chem}, it follows the set of evolution equations for chemical reactions:
\begin{equation*}
\begin{split}
\left\{ 
\begin{array}{l} 
\vspace{0.1cm}\\[-3mm]
\displaystyle\vspace{0.2cm}\;p _S= 0,\;\;\dot S = \lambda^{S},\;\;p_{N_{I}}=0,\;\; \dot{N}_{I}=\lambda^{N_{I}},\quad  I=1,...,R,\\
\displaystyle\vspace{0.2cm}\left(\dot{p}_{S}+ \frac{\partial U}{\partial S}\right)\left(\frac{1}{\frac{\partial U}{\partial S}}J_{a}\nu^{a}_{I} - \dot{N}_{I}\right)= 0,\quad I=1,...,R,\\
\displaystyle\vspace{0.2cm}\;\dot{p}_{N_{I}}-\dot{W}^{I}+ \frac{\partial U}{\partial N_{I}}= 0,\quad I=1,...,R,\\
\displaystyle\vspace{0.2cm} \;
\frac{\partial U}{\partial S}\dot S   = -J_a \nu^a_I \dot{W}^I.
\end{array} 
\right.
\end{split}
\end{equation*}
Using  $T= \frac{\partial U}{\partial S}$, $\mu^I=\dot W^I$ and $\mathcal{A} ^a= - \sum_{I=1} \nu^a _{I} \mu ^I$ and eliminating excess variables, we finally get the following evolution equations:
\begin{equation*}
\begin{split}
\left\{ 
\begin{array}{l} 
\vspace{0.1cm}\\[-3mm]
\displaystyle\vspace{0.2cm}\;\dot{N}_{I}=J_{a}\nu^{a}_{I},\quad I=1,...,R,\\
\displaystyle\vspace{0.2cm}\;\dot{W}^{I}= \frac{\partial U}{\partial N_{k}},\quad I=1,...,R,\\
\displaystyle\vspace{0.2cm} \;
T\dot S   = J_a \mathcal{A} ^a.
\end{array} 
\right.
\end{split}
\end{equation*}

\section{Conclusion}
In this paper, we have developed the Hamiltonian variational formulation for nonequilibrium thermodynamics, in which we have constructed the Hamilton-d'Alembert principle for nonlinear nonholonomic constraints of thermodynamic type and we have introduced a total  Hamiltonian by incorporating the primary constraints due to the degeneracy of a given Lagrangian. We have verified our theory with some illustrative examples of simple closed systems, i.e., thermodynamic systems with friction,  with internal matter transfer and with chemical reactions.

\paragraph{Acknowledgement.}
H.Y. is partially supported by Grants-in-Aid for Scientific Research (A) (17H01097),  Waseda University Research Projects (SR 2021C-134, SR 2021R-014, SR 2021C-518), JST CREST (JPMJCR1914), and the MEXT "Top Global University Project".


\end{document}